\documentclass[english,prl,twocolumn,notitlepage,superscriptaddress] {revtex4-1}

\pdfoutput=1

\usepackage{graphicx}
\usepackage{amssymb}
\usepackage{setspace}
\usepackage{epstopdf}
\usepackage{multirow}
\usepackage{bm}
\usepackage{braket}
\usepackage{amsmath}
\usepackage{makecell}
\usepackage{xcolor}

\usepackage{comment} 
\usepackage{graphicx}
\usepackage{color, colortbl}
\usepackage{ulem}
\newcommand\RED[1]{\textcolor{black}{#1}}

\begin{document}

\title{Multiplexed microwave resonators by frequency comb spectroscopy}

\author{Angelo Greco}
\email{angelo.greco@nano.cnr.it}
\affiliation{NEST, Istituto Nanoscienze-CNR and Scuola Normale Superiore, I-56127, Pisa, Italy}
\author{Jukka-Pekka Kaikkonen}
\affiliation{VTT, Technical Research Centre of Finland Ltd, QTF Centre of Excellence, P.O. Box 1000, FI-02044 VTT, Finland}
\author{Luca Chirolli}
\affiliation{Department of Physics and Astronomy, University of Florence, Sesto Fiorentino, Italy}
\author{Alberto Ronzani}
\affiliation{VTT, Technical Research Centre of Finland Ltd, QTF Centre of Excellence, P.O. Box 1000, FI-02044 VTT, Finland}
\author{Jorden Senior}
\affiliation{VTT, Technical Research Centre of Finland Ltd, QTF Centre of Excellence, P.O. Box 1000, FI-02044 VTT, Finland}
\author{Francesco Giazotto}
\affiliation{NEST, Istituto Nanoscienze-CNR and Scuola Normale Superiore, I-56127, Pisa, Italy}
\author{Alessandro Crippa}
\email{alessandro.crippa@cnr.it}
\affiliation{NEST, Istituto Nanoscienze-CNR and Scuola Normale Superiore, I-56127, Pisa, Italy}


\begin{abstract}
Coplanar waveguide resonators are central to the thriving field of circuit quantum electrodynamics. Recently, we have demonstrated the generation of a broadband microwave-frequency comb spectrum using a superconducting quantum interference device (SQUID) driven by a time-dependent magnetic field. Here, the frequency comb is used to spectroscopically probe a bank of coplanar microwave resonators, inductively coupled to a common transmission line, a standard circuit with a variety of applications. We compare the resonator line shape obtained from signals synthesized at room temperature using conventional electronics with the radiation produced in the cryogenic environment by our source, showing substantial equivalence in the estimation of the resonator quality factors. To measure non-uniformly spaced resonant frequencies, we drive the generator with a bi-chromatic tone to generate intermodulation products. Such a dense frequency comb spectrum enables simultaneous addressing of a few resonators via frequency multiplexing. Finally, we discuss the criteria for achieving effective spectroscopic coverage of a given frequency bandwidth.
\end{abstract}

\maketitle

\section{I.\,Introduction}
Superconducting coplanar microwave resonators have enabled the transposition of cavity quantum electrodynamics (QED) \cite{haroche2006} — focused on atom–photon interactions — to the mesoscopic realm, where artificial atoms couple to microwave photons \cite{blaisRMP}.
Today, they represent a widespread platform to investigate microscopic phenomena in optics and solid-state, with applications in material and device characterization \cite{ YacobycQED, phan2022detecting, buccheri2025microwave, Pop_loss}, quantum sensing \cite{sensing, kid}, electromagnetic environment engineering \cite{holst_environment, hofheinz_bright, cassidy_laser}, and quantum information processing \cite{blais2004cavity, goppl2008coplanar} where they accomplish coherent single-qubit control and state readout, allow strong qubit-photon coupling, and mediate qubit-qubit interactions \cite{wallraff2004strong, sagi2024gate, yu2023strong, petta_resonant}. 
One reason for the rapid growth of circuit QED (cQED) is the ability to generate microwave frequencies using mature commercial electronics to populate the resonators.\\ 
However, a central bottleneck in realizing large-scale quantum technology is that the number of interconnections and control lines is expected to scale with the number of sensors or qubits \cite{rent}, so both the physical size of the refrigerators and their cooling power must increase accordingly. 
Although frequency multiplexing partially alleviates this issue \cite{ustinov_multiplexing, multiplexed2012, multiplexedPhysRevApplied, multiplexing_sensors}, a substantial advance could come from the use of cryogenic-integrated control electronics in addition to the sensors array or the quantum processor \cite{pauka2021, cryogenicElectronics}. 
This would reduce the number of transmission lines from room temperature to the coldest stages while yielding minimal latency and low signal distortion. Amplifiers, switches, and signal synthesizers are only a few examples of a new class of electronic devices under development \cite{JAWS, tagliaferri2016modular, Mottonen2021, paghi}.\\
Recently, we have realized a cryogenic frequency comb emitter with very low power dissipation at cryogenic temperatures, a highly tunable driving signal, and a coherent broad-band output signal \cite{greco2025, bosisio2015parasitic, solinas2015, solinas2015radiation}. 
Here, we leverage the technique commonly known as frequency comb spectroscopy (FCS) \cite{reviewcomb_optics}, in which a frequency comb interacts with a target sample, to characterize the resonances of a bank of coplanar waveguide notch-type resonators. Our comb generator up-converts a pump frequency by synthesizing tens of harmonics whose frequencies are more than an order of magnitude higher than the pump frequency. The cryogenic generation of high-frequency tones a few centimeters from the target device significantly relaxes the constraints on the input wiring within the cryostat.
Our implementation of FCS relies on tuning the pump frequency, thereby shifting the harmonic mode that spectroscopically probes the resonator lineshape.
We compare the magnitude responses of the transmission parameters measured with a standard vector network analyzer (VNA) with the variations in the harmonic magnitude sensed via FCS.
Moreover, we show that a conventional comb structure with harmonics at integer multiples of the flux frequency is impractical for measuring resonance at non-uniformly spaced frequencies in the frequency domain. 
By driving the comb synthesizer with a bi-chromatic pump, we obtain a more complicated comb-like spectrum with many intermodulation products. 
We leverage some of these modes to simultaneously address multiple targets, thereby enabling multiplexed sensing of the resonator array. \RED{We conclude with a brief discussion on the theoretical optimization of the generated comb spectrum to efficiently cover a bandwidth of interest.}\\
The article is structured as follows. 
In Sect.\,II, we present the setup and describe our implementation of the frequency comb spectroscopy technique.
In Sect.\,III, we show the spectra of representative resonators obtained in two ways, namely by conventional input-output VNA measurements and by FCS.
In Sect.\,IV, we show that a bichromatic pump tone allows the measurement of a few non-uniformly spaced resonances via frequency multiplexing. \RED{Finally, in a perspective of reading out multiple qubits or arrays of sensors, we discuss the dense coverage of a given frequency bandwidth.}\\

\section{II.\, Device and frequency comb spectroscopy implementation}
\begin{figure}
    \centering
    \includegraphics[width=\columnwidth]{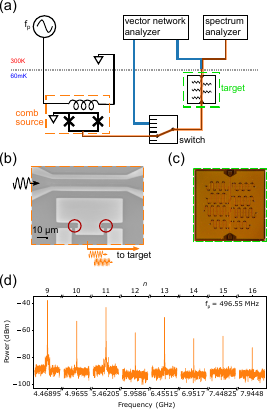}
    \caption{(a) Simplified diagram of the experimental setup. A pump tone at frequency $f_p$ modulates the flux threading a SQUID loop. The generated frequency comb is sent to the target chip, which hosts multiple resonators, via a electromechanical switch (orange path). The resonators can alternatively be measured using a VNA (blue path). (b) Comb generator. Scanning electron micrograph showing the dc SQUID and flux line. Dark gray indicates the superconducting Al, while the exposed areas of the substrate are light gray. The two red circles mark the Josephson junctions. The device produces microwave signals (orange arrows) whose frequencies are multiple integers of the pump tone (black arrow) applied via the flux line. (c) Target chip. Optical micrograph of the resonator chip, with the meanders inductively coupled to a common feedline. (d) Comb power spectrum. Harmonics in the 4-8\,GHz band generated by a pump tone of frequency 496.55\,MHz and amplitude 150\,mV at generator level. The reported power is measured by the spectrum analyzer at room temperature after a few stages of amplification. For details, see the complete circuit setup in Fig.\,\ref{fig:complete_setup}.}
    \label{fig:Fig1}
\end{figure}
Figure \,\ref{fig:Fig1}a reports a simplified schematic of the circuitry to measure the array of resonators.
The microwave frequency comb generator has been extensively outlined and characterized in Ref.\,\cite{greco2025}. Its core part is a dc superconducting quantum interference device (SQUID) realized by a standard shadow-mask technique in Al, which is shorted to ground on one side and connected to a coplanar waveguide on the other side (see Fig.\,\ref{fig:Fig1}b). 
The magnetic field threading the loop has a static component that sets the flux bias of the SQUID and a time-dependent component at frequency $f_p$, which induces rapid variations in the superconducting phase across the SQUID. 
Due to the ac Josephson effect, time-phase modulation leads to a train of voltage pulses with a repetition rate $f_p$. 
The emitted signal in the frequency domain corresponds to a frequency comb with $n$ modes, each of which has a frequency $f_n = n \times f_p$. The typical offset frequency of the comb spectra is zero here \cite{greco2025}.
This signal is routed out of the chip via a transmission line and then to the target chip, which hosts the resonators, via a low-loss coaxial cable.\\
As shown in Fig.\,\ref{fig:Fig1}a, we have interposed an RF switch between the comb generator and the resonators chip. 
This expedient allows us to interrogate the bank of resonators either with a signal from a VNA at room temperature (blue path) or with modes from our SQUID, the latter implementing an FCS scheme (orange path).\\
The target chip (see Fig.\,\ref{fig:Fig1}c) consists of a coplanar waveguide design where eight quarterwave resonators with resonant frequencies spanning the 4-8 GHz range are coupled to a 50 \,$\Omega$ transmission feedline. 
The structures are obtained from VTT’s fab process featuring sputtered Nb thin films on a high-resistance silicon substrate, patterned with 150\,mm wafer optical lithography and dry etching.
The resulting resonators routinely achieve internal quality factors in excess of one million at the single-photon level when optimally mounted in well-thermalized, infrared- and magnetic-shielded cryogenic environments, and have been recently employed in wafer-scale optical lithography fabrication of superconducting qubits \cite{VTT}.
\\
The comb source, the RF switch, and the resonators are anchored to the mixing chamber plate of a dilution refrigerator with a base temperature of 60 \,mK.
Figure \,\ref{fig:Fig1}d displays a typical comb spectrum generated by our source. We have chosen the pump frequency $f_p$ so that none of the harmonics intercepts any resonance. 
The power axis refers to the values sampled by the spectrum analyzer after the comb has passed through the amplification chain; see Appendix A for details. 
The signal emitted by the source that impinges on the target chip ranges between $-170$\,dBm and $-130$\,dBm, and is tuned by the amplitude of the flux modulation \cite{greco2025}. 
These numbers align with the amplitudes of other Josephson emitters \cite{peugeotPRApp} and fall within the typical values used in cQED experiments. 
The width of each frequency line is below our instrumental resolution, allowing us to plot the power spectrum rather than the power spectral density.\\

\section{III.\,Individual resonator spectroscopy}

\begin{figure*}[t]
\centering

\includegraphics[width=0.9\linewidth]{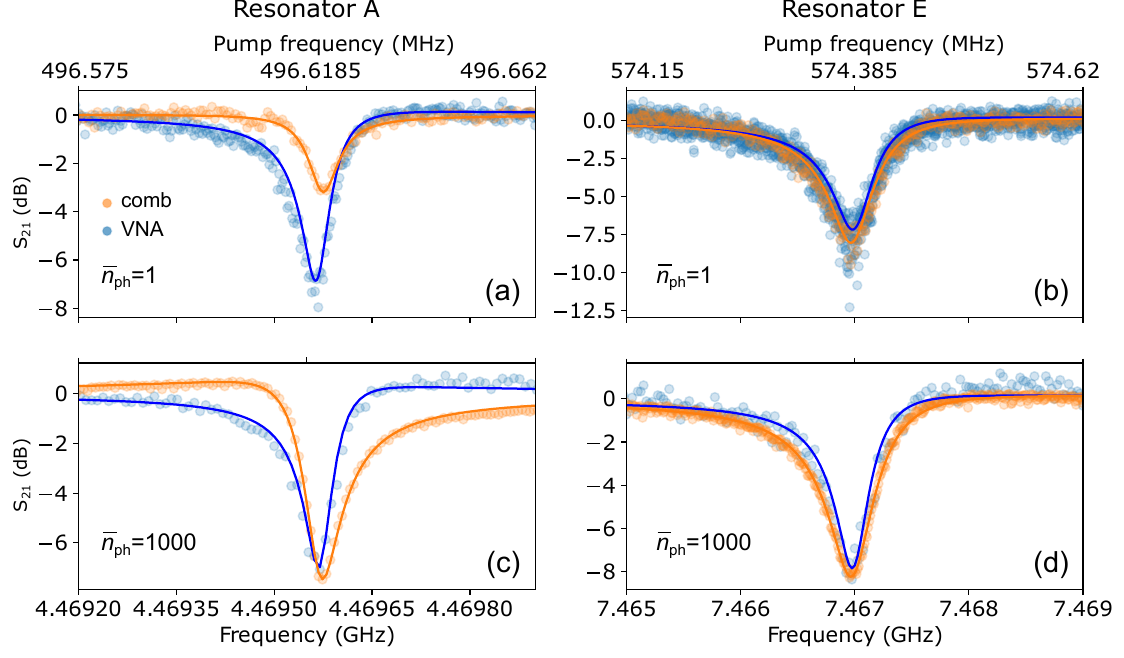}

\vspace{0.5cm}

\begin{minipage}{0.8\linewidth}
\centering
\begin{tabular}{|c|c|c|}
\hline
 & \makecell{Resonator A \\ $f^R_A=4.4696$ GHz} & \makecell{Resonator E \\ $f^R_E=7.4670$ GHz} \\
\hline
$\bar{n}_{\text{ph}}\approx1$ 
& \makecell{\textcolor{blue}{$Q_i=126$\,k, $Q_e=98$\,k, $\phi=0.28$}\\
          \textcolor{orange}{$Q_i=98$\,k, $Q_e=220$\,k, $\phi=-0.11$}}
& \makecell{\textcolor{blue}{$Q_i=27$\,k, $Q_e=19$\,k, $\phi=0.34$}\\
          \textcolor{orange}{$Q_i=32$\,k, $Q_e=21$\,k, $\phi=0.38$}} \\
\hline
$\bar{n}_{\text{ph}}\approx1000$ 
& \makecell{\textcolor{blue}{$Q_i=124$\,k, $Q_e=88$\,k, $\phi=0.43$}\\
          \textcolor{orange}{$Q_i=123$\,k, $Q_e=70$\,k, $\phi=-0.53$}}
& \makecell{\textcolor{blue}{$Q_i=25$\,k, $Q_e=15$\,k, $\phi=0.20$}\\
          \textcolor{orange}{$Q_i=25$\,k, $Q_e=15$\,k, $\phi=0.21$}} \\
\hline
\end{tabular}
\end{minipage}

\caption{Amplitude response of two exemplary resonators, A and E. Comparison of the frequency spectroscopy of resonator A obtained by standard input-output measurements by a VNA (light blue dots) or by FCS (orange dots) in the few-photon regime (a) and many-photon regime (c). The top x axes report the values of the pump frequency we swept in the FCS measurements, and the bottom x axes show the actual frequencies of the resonators spectra. The input power is calibrated to be the same for spectroscopy using the VNA and the frequency comb. (b), (d) Same plots but for resonator E. For FCS of resonator A we use the 9th harmonic, while for resonator E the 13th. The table at the bottom reports the fit parameters extracted from VNA and FCS data from panels (a) to (d). The unit of $\phi$ is rad.}
\label{fig:Fig2}

\end{figure*}

As a first step, we obtain the frequency profile of a resonator to compare the results obtained with the VNA and the FCS techniques. The resonance has to be measured with the same power in both cases, as the extraction of quality factors depends on the average photon occupation number $\bar{n}_{\text{ph}}$. 
The calibration of the input microwave power at the resonator is reported in Appendix B. We select the VNA or our comb generator as a microwave source by simply operating the cryogenic switch (Fig.\,\ref{fig:Fig1}a).\\
Comb spectroscopy is performed by sweeping the pump frequency $f_p$ over a range that allows the $n$-th harmonic to cross the resonance. In this way, the harmonic power varies according to the absorption of the resonator, as in standard $S_{21}(f)$ measurements by a VNA. The transmitted power is finally recorded by a spectrum analyzer, thus yielding the familiar resonator profile.\\
Figure \,\ref{fig:Fig2} shows two exemplary resonances among the five on the target chip (Appendix C for the full spectrum). 
The lineshape of resonators A and E, respectively the lowest and highest in frequency, is measured by  VNA or by FCS for $\bar{n}_{\text{ph}}\approx1$ [panels (a) and (b)], and for $\bar{n}_{\text{ph}}\approx1000$ [panels (c) and (d)], with $\bar{n}_{\text{ph}}$ the average photon occupation number. For FCS, the pump tone $f_p$ varies from 496.4\,MHz to 496.8\,MHz, so that the 9th harmonic covers the entire width of resonance A, according to $f_9=9\times f_p$. 
For resonator E, we sweep the pump from 574.15\,MHz to 564.72\,MHz using this time the 13th harmonic $f_{13}=13\times f_p$.\\
The VNA trace and the spectrum by FCS look different for Resonator A. In the few-photon regime (Fig.\,\ref{fig:Fig2}a), the depth of the resonance is more pronounced when sensed by the VNA ($-7$\,dB vs $-3$\,dB for FCS), while the resonant frequencies almost coincide. 
We extract the relevant parameters by fitting the experimental transmission data with the formula \cite{probst2015}
\begin{equation}\label{eq:S21}
    S_{21}(f)= 1- \frac{Q_i e^{-i\phi}}{\,Q_e + Q_i + 2 i Q_e Q_i \frac{f - f_0}{f_0}\,},
\end{equation}
where $Q_i$ denotes the internal quality factor, $Q_e$ the external quality factor, $f_0$ the resonance frequency, and $\phi$ quantifies the impedance mismatch \cite{probst2015} and the Fano interference effect \cite{rieger2023}. The fits of the VNA traces also account for the phase data of $S_{21}(f)$ (not shown), while by FCS we acquire only the amplitude component $|S_{21}(f)|$. We return to this aspect in the final section.\\
The different magnitudes of the dip leads to a discrepancy in the estimation of internal and external quality factors. However, such an inconsistency is lifted when many more photons populate the resonator, see Fig.\,\ref{fig:Fig2}c, where $\bar{n}_{\text{ph}}\approx1000$. 
The depth of the resonance is the same, but the shoulder of the dip is flipped, indicating a $\pi$ phase difference in the term $\phi$ of Eq.\,\ref{eq:S21}. 
The plots in panels (a) and (c) can be explained in terms of Fano physics by considering the different input circuitry when measuring with the VNA or by the radiation of our comb synthesizer. 
As argued in \,\cite{rieger2023}, an asymmetry factor $\phi$ emerges in transmission measurements whenever the field of an unwanted background path interferes with the field scattered by the resonator. Different input networks lead to variations in the relative phase and amplitude of the leakage path, thus modifying the resonator profile through different $\phi$ values and therefore changing the result of the fit, as reported in the table of Fig.\ref{fig:Fig2}.\\
For Resonator E, the traces acquired by the VNA and those obtained by FCS overlap well both in the few-photon regime (Fig.\,\ref{fig:Fig2}b) and in the many-photon condition (Fig.\,\ref{fig:Fig2}d). 
At these frequencies, Fano physics due to the different electrical path of the incoming signal has less impact. In either case, the quality factors $Q_i$ and $Q_e$ are comparable, thus validating the FCS as an alternative, less hardware demanding procedure for resonator spectroscopy.\\

\section{IV.\, Bi-chromatic pumping and frequency multiplexing}
\begin{figure}
\centering
\includegraphics[width=\columnwidth]{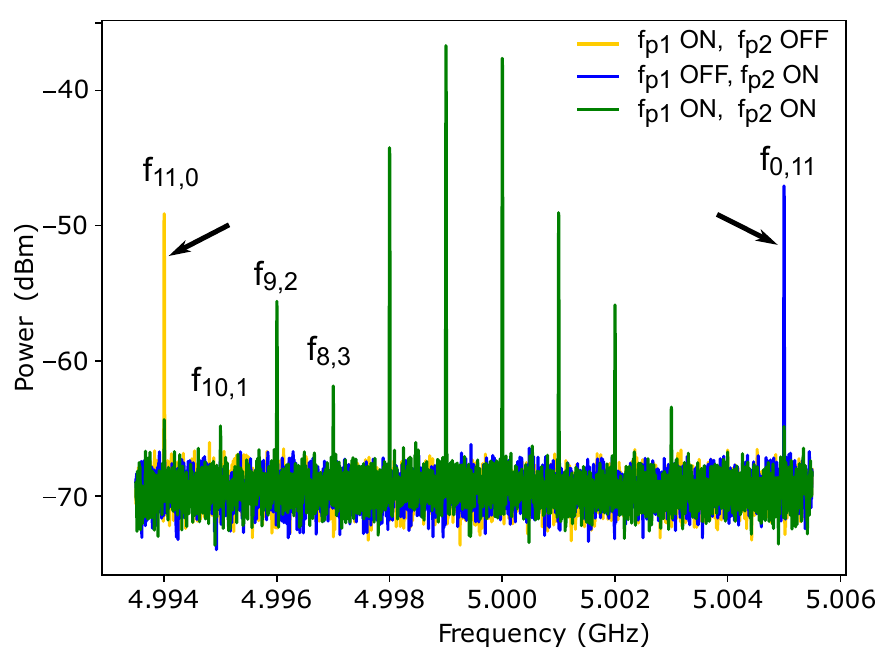} 
	\caption{Spectrum of a frequency comb originated by bi-chromatic pumping with $f_{p1}=454\,$MHz and $f_{p2}=455\,$MHz with amplitude 200\,mV at generator level. When one of the two tones is switched off, only the corresponding harmonics are present (blue and yellow spectra). When both tones are applied, the plot shows the intermodulation products (green data) generated around the midpoint of the 11th harmonics of the two pump tones. The subscripts indicate the pair of coefficients $n,m$ according to Eq.\,\ref{eq:intermodulation_products}.}
    	\label{fig:Fig3}
\end{figure}

\begin{figure*}
\centering
\includegraphics[width=\textwidth]{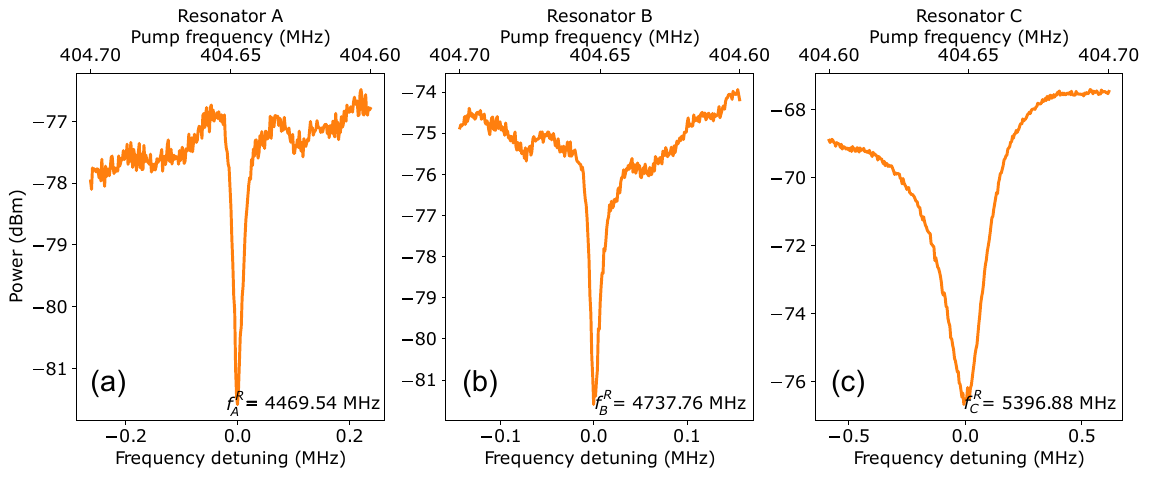} 
	\caption{Three resonators spectroscopy by frequency multiplexing. The pump frequencies are obtained from the system\,\ref{sist:multiplexing}: $f_{p1}=404.652465\,$MHz, $f_{p2}=541.065788\,$MHz. Only $f_{p1}$ is swept; see the top x-axis. $f_{p2}$ is idle. The coefficients to address resonators A, B and C (panels (a), (b), (c) respectively) are $(n_A,m_A) = (-5, 12)$, $(n_B,m_B) = (-3, 11)$ and $(n_C,m_C) = (12, 1)$. The bottom x axes indicate the resulting frequency detuning with respect to the resonance frequencies $f^R_A$, $f^R_B$ and $f^R_C$ reported in the plots.}
    	\label{fig:Fig4}
\end{figure*}
As detailed in our previous study\,\cite{greco2025}, the SQUID is a source of many harmonics of the pump frequency $f_p$. As such, frequency multiplexing naturally comes to mind, for instance, to address an array of sensors or to accomplish simultaneous few-qubit dispersive readout. However, the frequencies of our comb spectrum are constrained by the relation $f_n = n \times f_p$. As the resonant frequencies $f^R_i$ ($i=\text{A,B},\dots$ runs over the resonators) are non-uniformly spaced, finding a pump frequency $f_p$ such that $f^R_i= n_i \times f_p$, where $n_i \in \mathbb{N}$ and $f^R_i, f_p \in \mathbb{R}$, becomes in general complicated when the frequency spacing between $f^R_i$ gets narrow. In fact, this problem is equivalent to searching for a common divisor $f_p$ for all $f^R_i$.\\ 
For example, consider an extreme scenario with two non-overlapping resonances at $4.000\:$GHz and $4.001\:$GHz. To address both of them with a comb spectrum as discussed so far, we would need the 4000th and 4001st modes of a comb generated by a $1\:$MHz pump frequency. To our knowledge, no source capable of generating such an extended comb in the microwave range is available currently.\\
As a solution, we expand the configuration space by adding a second driving tone, so that the pump signal now consists of two frequencies, namely $f_{p1}$ and $f_{p2}$. Figure\,\ref{fig:Fig3}a shows a portion of the spectrum emitted by the SQUID when a bi-chromatic pump is fed into the flux line. The two black arrows indicate the harmonics of the two pumps ($11 \times f_{p1}$ and $11 \times f_{p2}$), recorded when only the related pump tone is active (yellow and blue spectra, respectively). Notably, additional, equally spaced lines with variable amplitude appear between them when both pumps are active (green trace).
Each of such intermodulation products has a frequency labeled with a pair of coefficients $(n,m)$ and satisfies the equation \cite{Root_Verspecht_Horn_Marcu_2013}:\\
\begin{equation}
    f_{n,m} = n\times f_{p1} + m \times f_{p2}.
\label{eq:intermodulation_products}
\end{equation}
It accounts for multi-photon mixing processes occurring when strong bi-chromatic tones couple to non-linear systems. Hence, the coefficients $n$ and $m$ can be positive or negative, provided that $f_{n, m}>0$. Equation\,\ref{eq:intermodulation_products} also yields the distance between consecutive intermodulation frequencies, which is the beating frequency $| f_{p1}-f_{p2}|$, equal to 1\,MHz in the present case. We can easily identify all the frequency lines in Fig.\,\ref{fig:Fig3}a accordingly. We note that the harmonics of the pump tones, i.e. $n \times f_{p1}$ and $m \times f_{p2}$, are more prominent when the relative pump is applied solely. 
At the same time, they are damped when a bi-chromatic drive is used, generating intermodulation products that distribute power across additional modes.\\
The overall envelope of such spectra is, in general, complicated, as it depends on the electromagnetic environment hosting the chip, the pump amplitude and frequency, the junction parameters, and the flux bias threading the SQUID loop.
An extended spectrum obtained under bi-chromatic pumping is provided in Fig.\,\ref{fig:double_pump_full_spectrum} in Appendix D.\\
As discussed above, the set of resonant frequencies $f^R_i$ can be obtained by combining the pump tones so that $f^R_i= n_i \times f_{p1} + m_i \times f_{p2}$, where $n_i, m_i \in \mathbb{Z}$. 
With this strategy, the multiplexing problem becomes tractable and the values of $f_{p1}$ and $f_{p2}$, together with the related coefficients $n_i$ and $m_i$, can be found numerically whenever an interval estimate is provided for each resonance. Experimentally, such tolerance can be of the same order as the linewidth of the resonance.\\
We now exploit the dense comb described in Eq.\,\ref{eq:intermodulation_products} by the bi-chromatic drive to address a few resonators at the same time by frequency multiplexing. In our target chip, we select three resonators (A, B, C) with resonant frequencies $f^R_A=4.46956\:$GHz, $f^R_B=4.73773\:$GHz, and $f^R_C=5.39690\:$GHz. To find suitable pump frequencies and mode numbers, we numerically solve the algebraic system for the three resonances:
\begin{equation}
\begin{cases}
f^R_A = n_A f_{p1} + m_A f_{p2}\\[4pt]
f^R_B = n_B f_{p1} + m_B f_{p2}\\[4pt]
f^R_C = n_C f_{p1} + m_C f_{p2}.
\end{cases}
\label{sist:multiplexing}
\end{equation}
The tolerance for the assigned resonance values is 50\,kHz. 
A solution of the system\,\ref{sist:multiplexing} is $f_{p1}=404.652465\:$MHz and $f_{p2}=541.065788\:$MHz, with $(n_A,m_A) = (-5, 12)$, $(n_B,m_B) = (-3, 11)$ and $(n_C,m_C) = (12, 1)$.\\
Figure\,\ref{fig:Fig4}a), b) and c) display the frequency profiles of the three resonators obtained by multiplexed FCS.
Measurement is performed by moving $f_{p1}$ around the calculated value with a span of 100 \,kHz. 
During this measurement, $f_{p2}$ is kept fixed. We acquire the power of the three intermodulation products of Eq.\,\ref{eq:intermodulation_products} at each step, thus obtaining the plots of panel (a), (b), and (c). The resonances are clearly resolved, and all resonant frequencies fall approximately at the same pump frequency, see the upper x-axis in each panel.
This confirms that the comb generated by the bi-chromatic drive has modes that can address at least three resonators simultaneously.\\
As a final note, it is worth investigating from a theoretical perspective how to choose the pumping tones to efficiently cover the frequency range where the resonators lie.  
We formally address the problem in Appendix E, and here we report and comment on the main conclusions.\\
Equation\,\ref{eq:intermodulation_products} defines a lattice of frequencies that is dense in $\mathbb{R}$ when the ratio of the two pump frequencies $f_{p1} / f_{p2}$ is irrational. 
We assume that a pump tone induces the generation of harmonics with a flat probability up to an index $\gtrsim 100$ \cite{greco2025}. Hence, the probability distribution of the sum of two independent harmonics (as given by Eq.\,\ref{eq:intermodulation_products}) is the convolution of their individual distributions. Finally, we refer to the probability of generating a frequency that matches an element of the lattice as density of states. It provides an estimate of how efficiently the generated comb spectrum covers the bandwidth of interest.\\
As we show in Appendix E, the form of such a density of states as a function of frequency resembles that of a low-pass filter, characterized by a flat region with high probability up to a cut-off frequency, after which the probability drops linearly. The width of the flat region depends on the values and the ratio of the two pump frequencies, which demonstrates a way to tailor the comb spectrum to maximize the overlap of the flat region with the bandwidth containing the frequencies to be addressed.\\
Further details and general statistical considerations about the lattice dimensions as a function of number of pump tones are reported in Appendix E.

\section{V.\, Conclusions}
In this work, we investigate superconducting coplanar waveguide resonators by using frequency comb spectroscopy. Our implementation relies on the cryogenic generation of a coherent microwave frequency comb by a highly tunable superconducting device (source) and a bank of notch-type resonators (target). By sweeping the pump tone of the source, we can reconstruct the frequency profile of any resonator in the bandwidth of our setup (4-8\,GHz). 
The fit of a few resonance lineshapes yields similar quality factors when comparing data by standard VNA characterization and by comb spectroscopy. 
Discrepancies in the resonance amplitudes and phase factors are attributed to different Fano interferences in the two measurement setups.
Next, we drive our cryogenic source with a bi-chromatic pump, which produces a frequency comb spectrum that includes many intermodulation products due to non-linearities in our SQUID. 
This solution relaxes the limitations on implementing frequency multiplexing using the modes of a standard comb. We engineer such a spectrum to implement multiplexed spectroscopy of three resonators and develop a model to further optimize the generated comb spectrum.\\
As a final note, we point out that we have not recorded the resonator phase swings by FCS because of instrumental limitations. 
Our pump generator imposes a random phase at the output tone each time we step the frequency. 
This randomizes the harmonic phase detected by the spectrum analyzer, since it is related to the pump phase \cite{greco2025}. As a matter of fact, it has prevented the reconstruction of the phase swing across the resonant frequency. 
Yet, we expect that advanced RF instruments (e.g., a nonlinear vector network analyzer, NVNA) capable of phase-synchronizing the output (pump) with multiple inputs (harmonics) can resolve the issue.


\section{ACKNOWLEDGMENTS}
The authors thank Dennis Rieger and Emanuele Enrico for helpful discussions. This research was supported by the PNRR MUR project PE0000023-NQSTI and by the European Union NextGenerationEU Mission 4 Component 1 CUP B53D23004030006 PRIN project 2022A8CJP3 (GAMESQUAD). We also acknowledge VTT and OtaNano Micronova cleanroom facilities, projects EU RIA No. 101007322 MatQu, No. 101113983 Qu-Pilot, Research Council of Finland through projects No. 356542 SUPSI, the QTF Centre of Excellence project No. 336817, and Technology Industries of Finland Centennial Foundation and Chips JU project Arctic No. 101139908. L.C. acknowledges the Fondazione Cariplo under the grant 2023-
2594. 

\section{APPENDIX A: CIRCUIT DIAGRAM}

\begin{figure*}
\centering
\includegraphics[width=\textwidth]{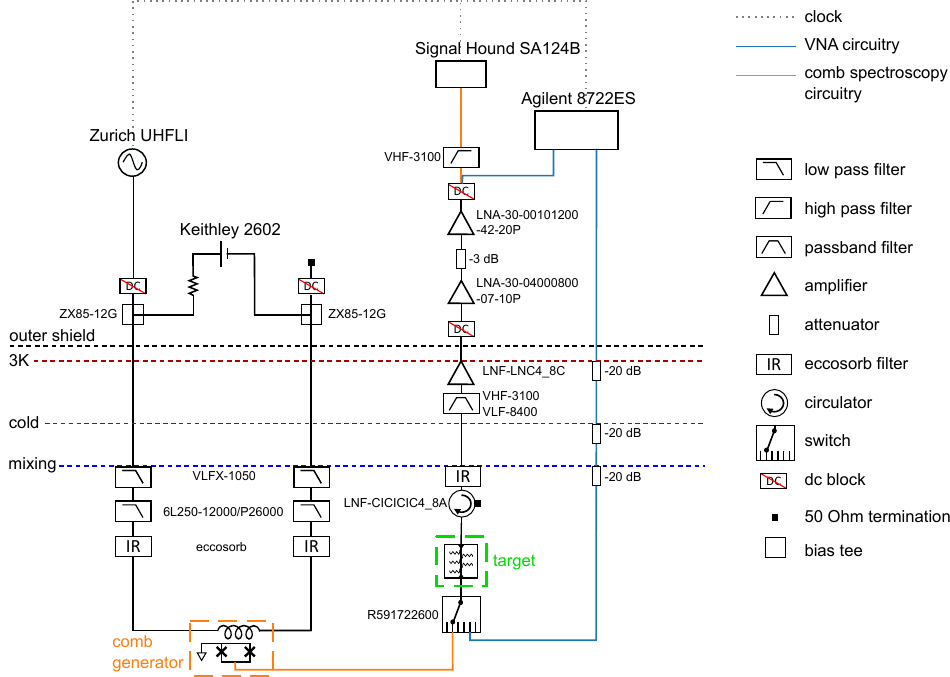} 
	\caption{Complete circuit diagram. The orange wires indicate the frequency comb spectroscopy setup-specific lines, and the light blue wires mark the VNA circuitry.}
    	\label{fig:complete_setup}
\end{figure*}

Figure\,\ref{fig:complete_setup} presents the complete circuit diagram used to perform the measurements discussed in the text. The room-temperature electronics comprises the RF source (Zurich UHFLI), a DC source to apply the flux bias to the SQUID (Keithley 2602), the spectrum analyzer for frequency comb spectroscopy (Signal Hound SA124B), and the vector network analyzer (Agilent 8722ES). All instruments are clock-synchronized.\\
In Fig.\,\ref{fig:complete_setup}, an important advantage of employing a cryogenic microwave source is worth noticing. The wiring for applying the time-dependent flux to the SQUID consists of compact coaxial cables designed for a dc-1 GHz bandwidth. From room temperature to the coldest plate, they attenuate 14-20\,dB within the 0.2-1\,GHz bandwidth. This circuitry is much less space-demanding than the standard CuNi coaxials used to perform the spectroscopy with the VNA.\\
To accurately rescale the measured power presented in Fig.\,\ref{fig:Fig1}d, Fig.\,\ref{fig:Fig3} and Fig.\,\ref{fig:double_pump_full_spectrum} to the power emitted at the comb generator output, it is necessary to quantify the total amplification along the readout chain. 
The total gain from the target output port to the data-acquisition instrumentation ranges from approximately 87 \,dB at 4 \,GHz to 82 \,dB at 8 \,GHz (hereafter concisely indicated as 87–82 \,dB). This gain includes 30–20\,dB of amplification within the cryostat output line over the same 4–8 GHz band—comprising the copper coaxial segment (orange colored) from the comb generator to the target sample and further on to the circulator, the TiN superconducting line from the circulator to the HEMT, the HEMT amplifier itself, and the CuNi coax leading to the output connector—as well as an additional two-stage room-temperature amplification of 28\,dB $-3$\,dB + 33\,dB.

\section{APPENDIX B: POWER CALIBRATION}
\label{sec:appendixA}

\begin{figure}
\centering
\includegraphics[width=\columnwidth]{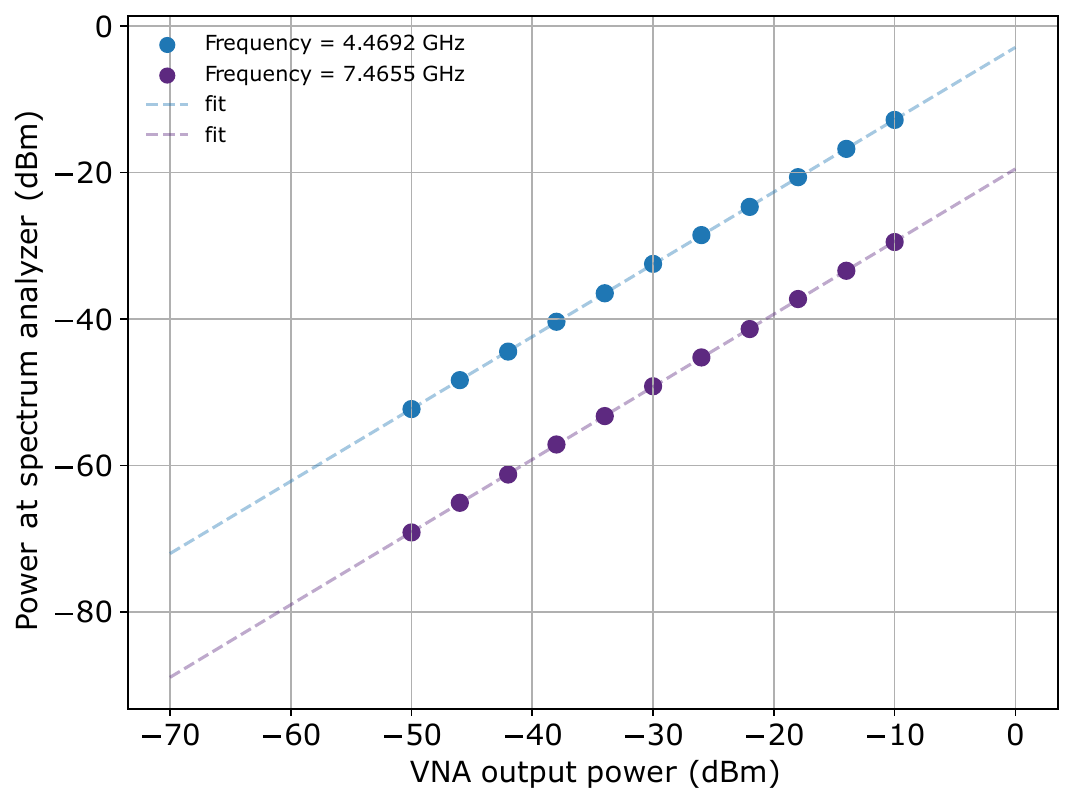} 
	\caption{Calibration of the transmission through the microwave setup. The power of a probing microwave signal generated by the VNA is measured by the spectrum analyzer as a function of the VNA output power. The data account for cable attenuation and amplification at cryogenic and room temperatures, as shown in Fig.\,\ref{fig:complete_setup}. Two frequencies close to the resonances of resonators A and E are chosen as representative of the attenuation in these portions of the spectrum. The linear fits yield coefficients of 1 and intercepts of $-2.91$\,dBm and $-19.50$\,dBm for the blue and purple data, respectively. This allows us to extrapolate the circuitry attenuation also into the single-photon regime, corresponding to nominal powers of $\sim-65$\,dBm at the lowest frequency and $\sim-50$\,dBm at the highest frequency.}
    	\label{fig:power_calibration}
\end{figure}
To properly compare the spectroscopy profiles of the resonators acquired by the VNA and FCS, the probing microwave power must be calibrated. The readout circuitry is common to both techniques, from the RF switch to the acquisition instrument at room temperature (see Fig.\,\ref{fig:complete_setup}). Hence, the equal power measured by the spectrum analyzer implies the same power reflected on the target chip. To tune the power of the comb modes, we need to vary the amplitude of the pump signal. 
However, it is more practical to calibrate the VNA power output. 
To do so, we set the RF switch to transmit the VNA output signal to the spectrum analyzer via the cryostat. The outcome is Fig.\ref{fig:power_calibration}. The intercepts of the straight lines yield the attenuation of the fridge circuitry at the two selected frequencies. With this graph in mind, we can adjust the VNA output power to ensure that the recorded powers from the comb generator and the VNA match. \\

\section{Appendix C: Full target resonators spectrum}
\begin{figure*}[h]
\centering
\includegraphics[width=\textwidth]{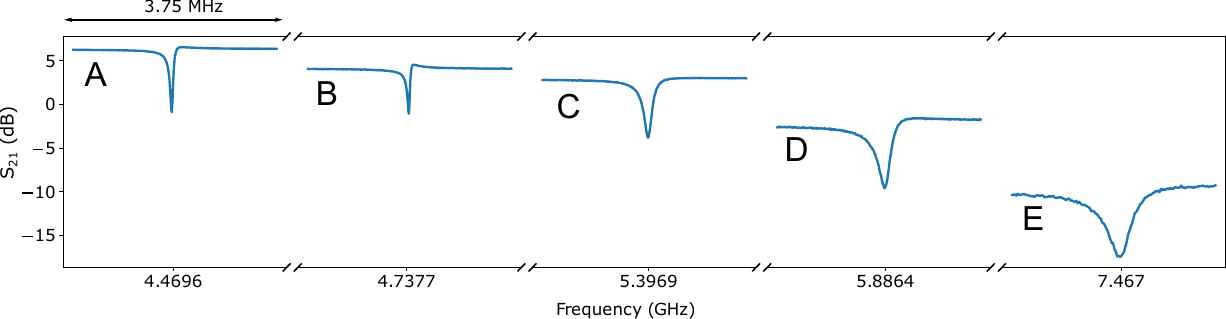} 
	\caption{Resonators spectra probed by the VNA. The output power at room temperature is set to $-10$\,dBm. We indicate each resonance frequency on the x-axis. The baseline of each spectral component decreases with increasing frequency due to damping, roughly linearly, as is typical for coaxial cables. Each plot covers a frequency span of 3.75\,MHz.}
    	\label{fig:chip_resonator_spectrum}
\end{figure*}
The comparative spectroscopy described in Section III is reported here for the other three resonators on the target chip. The results of the fit for internal and external quality factors and Fano factors ($Q_i, Q_e, \phi$, respectively) are reported in Table \,\ref{tab:quality_factors} for each resonance. Figure \ref{fig:chip_resonator_spectrum} shows the corresponding magnitude spectra measured by the VNA.

\begin{table*}[h] 
\vspace*{\fill}   
\centering
\begin{tabular}{|c|c|c|}
\hline
   & $\bar{n}_{\text{ph}}\approx1$ & $\bar{n}_{\text{ph}}\approx1000$ \\
\hline
\makecell{Resonator A \\ $f_r=4.4696$ } & \makecell{\textcolor{blue}{$Q_i=126$, $Q_e=98$, $\phi=0.28$} \\ {\textcolor{orange}{$Q_i=98$, $Q_e=220$, $\phi=-0.11$}}}
            & \makecell{\textcolor{blue}{$Q_i=124$, $Q_e=88$, $\phi=0.43$} \\ {\textcolor{orange}{$Q_i=123$, $Q_e=70$, $\phi=-0.53$}}} \\
\hline
\makecell{Resonator B \\ $f_r=4.7377$ } & \makecell{\textcolor{blue}{$Q_i=148$, $Q_e=164$, $\phi=0.58$} \\ {\textcolor{orange}{$Q_i=120$, $Q_e=99$, $\phi=0.28$}}}
            & \makecell{\textcolor{blue}{$Q_i=144$, $Q_e=143$, $\phi=0.70$} \\ {\textcolor{orange}{$Q_i=129$, $Q_e=82$, $\phi=-0.32$}}} \\
\hline
\makecell{Resonator C \\ $f_r=5.3967$ } & \makecell{\textcolor{blue}{$Q_i=53$, $Q_e=48$, $\phi=0.24$} \\ {\textcolor{orange}{$Q_i=67$, $Q_e=31$, $\phi=0.52$}}}
            & \makecell{\textcolor{blue}{$Q_i=54$, $Q_e=47$, $\phi=0.16$} \\ {\textcolor{orange}{$Q_i=55$, $Q_e=29$, $\phi=0.09$}}} \\
\hline
\makecell{Resonator D \\ $f_r=5.8864$ } & \makecell{\textcolor{blue}{$Q_i=36$, $Q_e=23$, $\phi=0.50$}\\ {\textcolor{orange}{$Q_i=47$, $Q_e=10$, $\phi=0.53$}}}
            & \makecell{\textcolor{blue}{$Q_i=46$, $Q_e=27$, $\phi=0.59$} \\ {\textcolor{orange}{$Q_i=42$, $Q_e=10$, $\phi=0.47$}}} \\
\hline
\makecell{Resonator E \\ $f_r=7.4670$ } & \makecell{\textcolor{blue}{$Q_i=27$, $Q_e=19$, $\phi=0.34$} \\ {\textcolor{orange}{$Q_i=32$, $Q_e=21$, $\phi=0.38$}}}
            & \makecell{\textcolor{blue}{$Q_i=25$, $Q_e=15$, $\phi=0.20$} \\ {\textcolor{orange}{$Q_i=25$, $Q_e=15$, $\phi=0.21$}}} \\
\hline
\end{tabular}
\caption{
Fit parameters extracted from fits using VNA and FCS techniques. 
Blue lines refer to VNA, while orange ones refer to FCS. All the fits are performed at average occupation numbers $\bar{n}_{\text{ph}}\approx1$ and $\bar{n}_{\text{ph}}\approx1000$.\\
The units in the table are: frequency = [GHz], quality factor = [$10^3$], phase = [rad].
}\label{tab:quality_factors}
\vspace*{\fill}  
\end{table*}

\section{Appendix D: Full bi-chromatic comb spectrum}
\begin{figure*}[h]
\centering
\includegraphics[width=\textwidth]{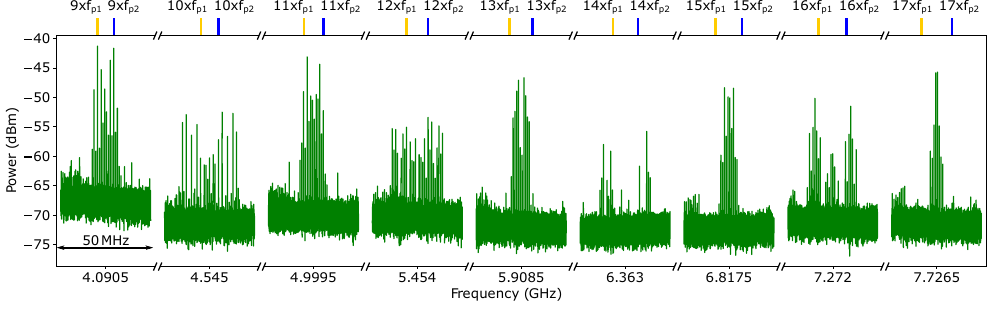} 
	\caption{Bi-chromatic comb spectrum. Patch of nine spectra with bi-chromatic pumping with $f_{p1}=454\,$MHz and $f_{p2}=455\,$MHz, each with an amplitude of 200\,mV at the RF source. On the top x-axis, the frequency position of the harmonics of the two pump tones is marked. Yellow ticks indicate the harmonics of $f_{p1}$, while blue ticks point to the harmonics of $f_{p2}$. The frequency span of each dataset is 50\,MHz.}
    	\label{fig:double_pump_full_spectrum}
\end{figure*}
Figure \ref{fig:double_pump_full_spectrum} shows the frequency comb spectrum generated by a bi-chromatic drive in the $4-8\:$GHz range. The pump frequencies are $f_{p1}=454\,$MHz and $f_{p2}=455\,$MHz, as in Fig.\,\ref{fig:Fig3}. 
The harmonics of either pump tone are marked by the yellow and blue ticks on the top x-axis. The spectrum presents nucleations of intermodulation products both between and outside the pump harmonics, since both positive and negative coefficients are allowed in the mixing processes (recall Eq.\,\ref{eq:intermodulation_products}) known to appear when dealing with nonlinear systems driven by two strong monochromatic tones \cite{Root_Verspecht_Horn_Marcu_2013}.

\section{Appendix E: Matching a set of target frequencies with a frequency comb}

\subsection{Statement of the problem}

We have a set of $N$ frequencies $\omega_i$, for $i=1,\ldots,N$, randomly distributed in a bandwidth $(W_{\rm min},W_{\rm max})$, with $BW=W_{\rm max}-W_{\rm min}$, so that $W_{\rm min}<\omega_i<W_{\rm max}$. Recalling Eq.\,\ref{eq:intermodulation_products}, we want to find two frequencies $f_{p1}$ and $f_{p2}$ that satisfy
\begin{equation}
m_{i1}f_{p1}+m_{i2}f_{p2}=\omega_i.
\end{equation}
The system describes $N$ non-linear equations for $2N+2$ variables and the problem is highly undetermined. If we relax the equality, we can pose the problem of finding the best values of $f_{p1}$ and $f_{p2}$ for which we have
\begin{equation}
m_{i1}f_{p1}+m_{i2}f_{p2}\simeq \omega_i.
\end{equation}
Indeed, for $f_{p1}/f_{p2}$ irrational, the lattice $\Lambda=\{m_1f_{p1}+m_2f_{p2} ~ | ~ (m_1,m_2)\in \mathbb{Z}^2\}$ is dense in $\mathbb{R}$, and the comb can get close enough to the target frequencies to match them within an error, which can eventually be supplied by the linewidth of the resonances.\\
We first consider the general case of a set of $N$ real numbers representing the target frequencies centered around zero
\begin{equation}
\{\omega_i\}_{i=1}^N \subset [-W/2,W/2],
\end{equation}
and consider their approximation by integer linear combinations of $K$ pump tones
\begin{equation}
\omega_i \approx \sum_{j=1}^K m_{ij} f_{pj},
\qquad m_{ij} \in \mathbb{Z},
\end{equation}
where the frequencies $\{f_{pj}\}$ are assumed to be rationally independent. In this case,  the projected lattice reads $\Lambda=\{\sum_{i=1}^Km_if_{pi} ~ | ~ m_i\in \mathbb{Z}\}\subset \mathbb{R}$, is dense in $\mathbb{R}$, the larger is $K$, the denser is the set in $\mathbb{R}$. Restricting the integers to $|m_j| \le M$, the number of distinct lattice points scales as
\begin{equation}
\#\Lambda_M \sim M^K,
\end{equation}
that is the volume in the $K$-dimensional space, while their projections fill an interval of width $\sim M$.
Therefore, the mean spacing between lattice points scales as
\begin{equation}
\Delta(M) \sim \frac{1}{M^{K-1}}.
\end{equation}
To optimally approximate $N$ target frequencies in a bandwidth $W$, the lattice spacing should match the mean level spacing,
\begin{equation}
\Delta_{\mathrm{opt}} \sim \frac{W}{N}.
\end{equation}
so that the effective integer cutoff is $M \sim N^{1/(K-1)}$. Substituting back, the minimal achievable spacing becomes
\begin{equation}
\Delta_{\mathrm{opt}} \sim \frac{W}{N^{K-1}}.
\end{equation}
Furthermore, defining $\Pi_{\Lambda}(\omega_i)$ the nearest lattice point to $\omega_i$, the quantization error
\begin{equation}
\varepsilon = \omega - \Pi_{\Lambda}(\omega).
\end{equation}
 is uniformly distributed in the interval $\varepsilon \in \left[-\frac{\Delta}{2}, \frac{\Delta}{2}\right]$ and therefore the mean square error (MSE) is \cite{GrayNeuhoff1998}
\begin{equation}
\mathrm{MSE}
= \mathbb{E}[\varepsilon^2]
= \frac{1}{\Delta} \int_{-\Delta/2}^{\Delta/2} \varepsilon^2\, d\varepsilon
= \frac{\Delta^2}{12}.
\end{equation}
Using the optimal spacing yields the scaling
\begin{equation}
\mathrm{MSE}_{\mathrm{opt}}
\sim
\frac{W^2}{12\, N^{2(K-1)}}.
\end{equation}
As intuition suggests, the higher the number $K$ of pump frequencies, the smaller the level spacing of the comb, so a better matching is possible. 
At the same time, the integer $M$ is set by the number of harmonics that the pump naturally generates, so we can consider it to be more or less fixed by the device and circuitry, and typically in a problem of interest we have $M\gg N$, with $N\sim 10\div 20$ and $M\sim 500\div 1000$. It then follows that increasing the number of pump frequencies does not yield statistically better results, so in the following we restrict ourselves to the case of two pump frequencies.

\subsection{Comb density of states}
\begin{figure*}[h]
\centering
\includegraphics[width=0.8\textwidth]{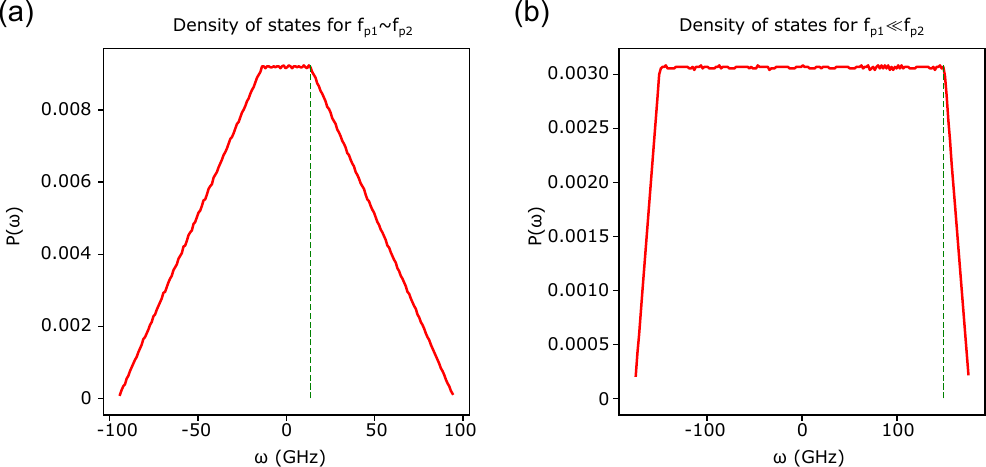} 
	\caption{Simulated density of states of the generated frequencies for two representative cases: in (a) $f_{p1}=404.652465$\,MHz and $f_{p2}=541.065788$\,MHz, in (b) $f_{p1}=404.652465/3$\,MHz and $f_{p2}=3\times 541.065788$\,MHz. The dashed vertical green line is at the frequency $M|f_{p1}-f_{p2}|$, assuming $M=100$.}
    	\label{fig:dos}
\end{figure*}

We now examine the density of states of the lattice generated by two pump frequencies. The frequency comb that is generated by the pump tone is distributed around zero frequency, i.e. there are both positive and negative harmonics. The mean level spacing of the comb is $\Delta=(f_{\rm p, max}-f_{\rm p, min})/\#\Lambda$ and it reads 
\begin{equation}
\Delta=\frac{2M}{(2M+2)^2}(f_{p1}+f_{p2})\sim\frac{f_{p1}+f_{p2}}{2M},
\end{equation}
where $f_{\rm p, max}=-f_{\rm p, min}=M(f_{p1}+f_{p2})$ and $\#\Lambda=(2M+2)^2$. The distribution of the sum of two independent variables is given by the convolution of their distribution. Each frequency is generated with a flat probability
\begin{equation}
P_n=\Theta(M-|n|)/(2M+1),
\end{equation}
with $\Theta(x)$ the Heaviside function. Since the linear combination $f_{p1}n+f_{p2}m$ is a real number, it is convenient to express the distribution of each frequency as $P_i(x)=\sum_{n}P_n\delta(x-f_{pi}n)$, so that the convolution of the two probabilities is
\begin{equation}
P(\omega)=\sum_{n,m=-M}^MP_nP_m\delta(\omega-f_{p1}n-f_{p2}m).
\end{equation}
The $\delta$-function allows us to account for a line broadening $\gamma$ that is naturally present in the target resonator linewidth, so that it can be replaced with a  simple step function of width $\gamma$ and height $1/\gamma$, so that we can approximately write
\begin{equation}
P(\omega)=\frac{1}{\gamma(2M+1)^2}\sum_{n=-M}^M\Theta(M-|n|)\Theta(M-[|\omega-f_{p1}n|/f_{p2}]),
\end{equation}
with $[\cdot]$ the integer part, which gives the probability of matching the frequency $\omega$ with the comb frequency within an error $\gamma$.\\
Taking into account both positive and negative frequencies, the resulting density of states is trapezoidal, with a flat central region of equal probability $2/(M(3|f_{p1}-f_{p2}|+f_{p1}+f_{p2}))$ in the region $|\omega|<M|f_{p1}-f_{p2}|$ and linearly decreasing boundary zones. The relative dominance of the two regions is determined by the ratio $f_{p1}/f_{p2}$: the closer $f_{p1}/f_{p2}$ to 1, the smaller becomes the flat central region of equal probability and the larger/smaller is $f_{p1}/f_{p2}$ than 1, the wider is the central flat region. This is understood considering that in the limiting case $f_{p1}=f_{p2}$ the distribution is pyramidal, with a triangular DoS, and that the distribution is cut to a trapezoidal shape when $f_{p1}\neq f_{p2}$. In Fig.~\ref{fig:dos} we show the DoS relative to two cases: i)  $f_{p1}=404.652465$ MHz and $f_{p2}=541.065788$ MHz; ii) $f_{p1}=404.652465/3$ MHz and $f_{p2}=3\times541.065788$ MHz.\\
Typically, the target frequencies that we want to approximately match belong to a set that is shifted from zero, $W_{\rm min}<\omega_i<W_{\rm max}$. The strategy for choosing the pump frequencies $f_{p1}$ and $f_{p2}$ depends on the experimental possibility of moving to the rotating frame at $(W_{\rm min}+W_{\rm max})/2$. If this is possible, then the best choice is to choose similar pump frequencies, $f_{p1}\sim f_{p2}$, while maintaining the constraint of irrational $f_{p1}/f_{p2}$. In turn, if experimentally there is no way to effectively apply a shift to the target frequencies, we can choose the two pump frequencies such that $f_{p1}\ll f_{p2}$, so we effectively move to the rotating frame with the higher frequency and fine tune with the lower one. In this case, the optimal choice is to set the larger frequency to match the mean level spacing $\Delta$, and the other pump frequency one order of magnitude lower. 


%

\end{document}